\documentclass[aps,prl,a4paper,twocolumn,longbibliography,superscriptaddress]{revtex4-1}
\usepackage{amsmath}
\usepackage{amssymb}
\usepackage{mathtools}
\usepackage{graphics}
\usepackage{comment}
\usepackage{tikz}
\usepackage[normalem]{ulem}
\usepackage{soul}
\usepackage{bm}

\newcommand{\beq}{\begin{equation}}
\newcommand{\eeq}{\end{equation}}
\newcommand{\bei}{\begin{itemize}}
\newcommand{\eei}{\end{itemize}}
\newcommand{\ben}{\begin{enumerate}}
\newcommand{\een}{\end{enumerate}}

\usepackage{hyperref}
\hypersetup{
   pdfpagemode=None, 
   pdfstartpage=1,
   pdfmenubar=true,
   pdftoolbar=true,
   colorlinks = true,
   linkcolor=blue,
   citecolor=blue,
   urlcolor=blue,
   bookmarksopen=false
 }

\usepackage[normalem]{ulem}

\definecolor{darkblue}{rgb}{0.,0.24,0.51}
\definecolor{britishracinggreen}{rgb}{0.0, 0.26, 0.15}
\definecolor{darkgreen}{rgb}{0,0.60,.2}

\begin{document}
\title{Probing Superfluidity with Quantum Vortex Necklaces}
\author{Andrea Richaud}
\affiliation{Departament de F\'isica, Universitat Polit\`ecnica de Catalunya, Campus Nord B4-B5, E-08034 Barcelona, Spain}
\author{Pietro Massignan}
\affiliation{Departament de F\'isica, Universitat Polit\`ecnica de Catalunya, Campus Nord B4-B5, E-08034 Barcelona, Spain}
\date{\today}
	
\begin{abstract}
We present a method for measuring the superfluid fraction of a Bose-Einstein condensate (BEC) without relying on external perturbations or imposed optical lattices. Our approach leverages the intrinsic rotation of vortex necklaces in one component of a binary superfluid mixture, where the vortex cores act as effective potential wells for the second component. The rotation of the vortex necklace transfers angular momentum to the latter, enabling a direct determination of its effective moment of inertia. Comparing this value with its classical counterpart allows us to extract the superfluid fraction, which we find to be precisely bracketed by the Leggett bounds. By increasing intercomponent interactions, the second component undergoes a crossover from a delocalized and fully superfluid state to an insulating state consisting of a regular array of localized density peaks. Furthermore, the dynamical instability of vortex necklaces provides a natural framework for investigating superfluidity in dynamically evolving and disordered landscapes.
\end{abstract}
\maketitle

\textit{Introduction. ---} 
Bose-Einstein condensation and superfluidity, while often co-occurring, are fundamentally different phenomena. The former is a static property quantifying the macroscopic occupation of a single quantum state and is represented by a scalar, while the latter is a dynamical quantity measuring the (absence of) response to a weak and constant perturbation and is therefore a tensor~\cite{Stringari2016}. In weakly interacting uniform ultracold bosonic systems the condensed and superfluid fractions typically coincide, but they can be significantly different in other situations: an ideal BEC, for example, is fully condensed but not superfluid, whereas superfluid liquid helium at zero temperature exhibits complete superfluidity despite having only $\sim 10\%$ condensate fraction~\cite{Leggett1999}. 

A foundational approach to probing the superfluid fraction originated with the Andronikashvili experiment~\cite{Andronikashvili1971}, which measured the frequency shift of a torsional pendulum immersed in liquid helium. As temperature decreases, the emergence of a superfluid component, which decouples from the pendulum's motion, leads to a measurable increase in oscillation frequency, reflecting a reduced effective moment of inertia. This pioneering method inspired a broad family of moment-of-inertia-based techniques~\cite{Stringari1996,Leggett1999,Nyeki2017} for the determination of the superfluid fraction, including recent applications to toroidal dipolar supersolids~\cite{Tanzi2021,Tengstrand2021,Roccuzzo2022,Hertkorn2024,Sindik2024}. Other experiments extracted the superfluid fraction by measuring anisotropy of sound propagation~\cite{Chauveau2023}, scissor modes~\cite{Tao2023}, and transport under lattice modulations~\cite{Pezze2024, Biagioni2024}.

The idea of probing conduction via rotation, or equivalently through a synthetic gauge field, also arises in Fermi lattice models through the concept of the Drude weight~\cite{Giamarchi1995}, which quantifies the singular part (at zero frequency) of the electrical conductivity. Originally introduced to distinguish metals from insulators~\cite{Kohn1964}, it has recently been applied to characterize transport in ring-like multicomponent Fermi systems~\cite{Richaud2021PRB}, minimal quantum circuits capable of sustaining persistent currents~\cite{Polo2024}.

\begin{figure}[b]
    \centering
    \includegraphics[width=\columnwidth]{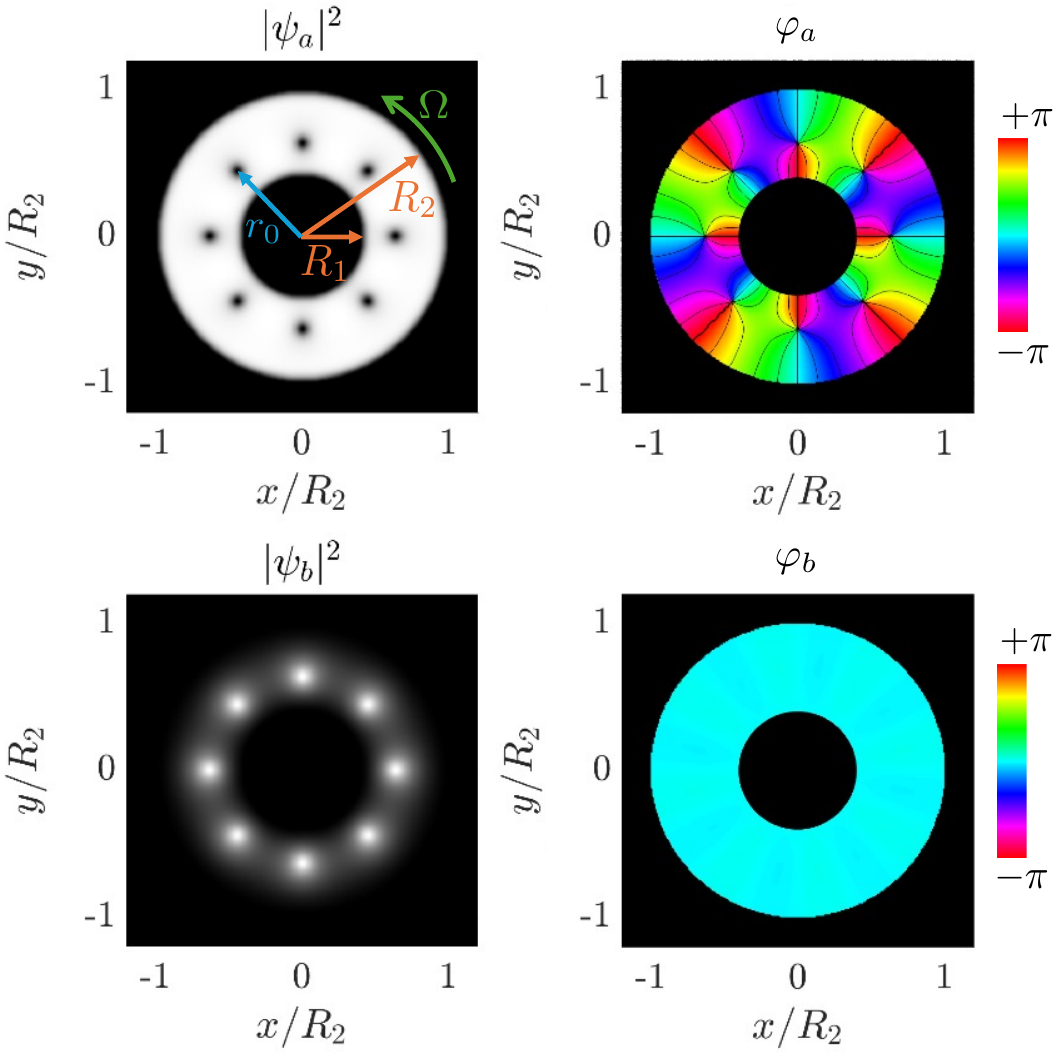}
    \caption{\textbf{Upper row:} density (left) and phase (right) of the majority component $a$, hosting a necklace with $N_v=8$ vortices, which rotates at frequency $\Omega$. \textbf{Lower row:} density (left) and phase (right) of the minority component $b$. Parameters are given in the text; here $g_{ab}/\sqrt{g_{aa}g_{bb}}=0.4$.}
    \label{fig:Introductory_figure}
\end{figure}

In this work, we propose a conceptually different strategy: introducing an auxiliary BEC component that acts as a built-in probe to assess the superfluid properties of the condensate under test. Specifically, we investigate a binary BEC mixture, where one component hosts a rotating necklace of quantized vortices, acting as a periodic, angularly modulated potential for the second component (see Fig.~\ref{fig:Introductory_figure}). The intrinsic rotation of this vortex array enables angular momentum transfer to the second component, thus supporting a direct, non-invasive measurement of the second component’s moment of inertia and, consequently, its superfluid fraction. This system circumvents the need for artificial lattices~\cite{Chauveau2023,Tao2023} or time-dependent forcing~\cite{Sindik2024}, while naturally allowing for the exploration of the crossover from the fully superfluid to the localized (Mott-like~\cite{Fisher1989,Jaksch1998,Greiner2002}) phase thanks to the tunability of the intercomponent interaction. Moreover, the same platform inherently lends itself to the study of out-of-equilibrium phenomena, such as disordering or slow ramps across the miscibility threshold, making it ideal for investigating dynamical aspects of superfluidity. The annular geometry we consider is particularly suited to such investigation, in light of recent experimental advances in annular trapping and vortex control~\cite{Kwon2021,Simjanovski2023,Hernandez2024,Pezze2024,Simjanovski2025,Grani2025}.

\smallskip
\textit{Rotating vortex necklaces. ---} We consider a binary mixture of atomic Bose-Einstein condensates consisting of $N_a$ atoms of component $a$ and $N_b$ atoms of component $b$ with atomic masses $m_a$ and $m_b$. We assume $N_a \gg N_b$, so that we can refer to $a$ ($b$) as the majority (minority) component. The system is subject to a strong harmonic confinement along the $z$-axis, associated to the harmonic-oscillator length $d_z$, which effectively freezes atomic motion in this direction. In the $xy$ plane, the mixture is trapped in an annular confining potential $V_\mathrm{tr}$ with hard walls at inner and outer radii $R_1$ and $R_2$ (see Fig.~\ref{fig:Introductory_figure}).
At zero temperature, the state of the system is described by two order parameters $\psi_a$ and $\psi_b$ (with normalization $\int |\psi_i|^2\,\mathrm{d}^2 r =N_i$) whose time evolution is governed by the coupled Gross-Pitaevskii equations (GPEs)
\begin{equation}
    \label{eq:GPEs}
    i\hbar\frac{\partial \psi_i}{\partial t} =\left(-\frac{\hbar^2\nabla^2}{2m_i}+V_\mathrm{tr}+ \sum_{j=a,b} g_{ij} |\psi_j|^2\right)\psi_i,\quad i=a,b
\end{equation}
where the coupling constants $g_{ij}=\sqrt{2\pi} \hbar^2 a_{ij}/(m_{ij}d_z)$ depend on the intra- and intercomponent $s$-wave scattering lengths $a_{ij}$, as well as on the reduced masses $m_{ij}=\left(m_i^{-1}+ m_j^{-1}\right)^{-1}$~\cite{Hadzibabic2011,Barenghi2016}.

The GPE for component $a$ admits metastable solutions featuring $N_v$ quantized vortices arranged as a regular polygon. These “vortex necklaces” have been extensively studied in the literature, as they appear in both classical inviscid fluids~\cite{Thomson1871,Havelock1931,Mertz1978} and rotating superfluids~\cite{Fetter1967,Kim2004,Caldara2023,Chaika2023,Hernandez2024,Caldara2024}. Similar geometric patterns emerge also in charged plasmas, where they are referred to as ``Wigner crystals" (see Ref.~\cite{Neely2024} and references therein).  

For component $b$, in the absence of intercomponent interactions  ($g_{ab}=0$), the density distribution $|\psi_b|^2$ is approximately uniform within the annulus. When a repulsive intercomponent interaction is introduced ($g_{ab}>0$), the vortex necklace in the majority component acts as an effective optical lattice along the azimuthal direction $\theta$, with vortex cores serving as potential wells for the minority component.

A crucial property of this system, which we will extensively exploit, is that vortex necklaces are not stationary in the laboratory frame, but instead rotate rigidly due to mutual interactions between vortices and the presence of boundaries. For vortex necklaces of radius $r_0=\sqrt{R_1R_2}$, the precession frequency $\Omega$ simply reads
\begin{equation}
    \label{eq:Omega}\Omega\left(r_0=\sqrt{R_1R_2}\right) =\frac{\hbar}{m_a r_0^2}\left(\mathfrak{n}_1 +\frac{N_v-1}{2}\right)
\end{equation}
with $\mathfrak{n}_1\in\mathbb{Z}$ the number of quanta of circulation around the inner boundary of the annulus~\cite{Caldara2024}.

\smallskip
\textit{Probing the superfluid fraction. ---} 
From the perspective of component-$b$ atoms, component $a$ acts as a \emph{rotating} lattice potential. That means that the vortex necklace in component $a$ not only imparts a density modulation on $\psi_b$, but also a non-zero angular momentum. As noted above, a rotating potential provides an ideal framework for measuring the superfluid fraction of a quantum fluid. In this sense, the majority component $a$ serves as a natural probe for assessing the superfluid properties of the minority component $b$. In what follows, we focus on the diagonal azimuthal component of the superfluid tensor, $f_{\theta\theta}$, which quantifies the azimuthal response of component $b$ to an azimuthal stimulus. This scalar, referred to as the ``superfluid fraction", is defined as
\begin{equation}
     f_{\theta\theta} = 1- \lim_{\Omega\to 0} \frac{\langle\psi_b|\hat{L}_z|\psi_b\rangle}{\Omega \int m_b|\psi_b|^2 r^2\,\mathrm{d}^2 r },
     \label{eq:f_thetatheta}
\end{equation}
where $\hat{L}_z=-i\hbar\partial_\theta$ the third component of the angular-momentum operator. The value of $f_{\theta\theta}$ lies between 0 and 1 and captures the deviation of the system’s actual angular momentum $\langle\psi_b|\hat{L}_z|\psi_b\rangle$ from its classical rigid-body counterpart $\Omega \int m_b|\psi_b|^2 r^2\,\mathrm{d}^2 r$~\cite{Stringari1996}.

\begin{figure}[b]
    \centering
    \includegraphics[width=\columnwidth]{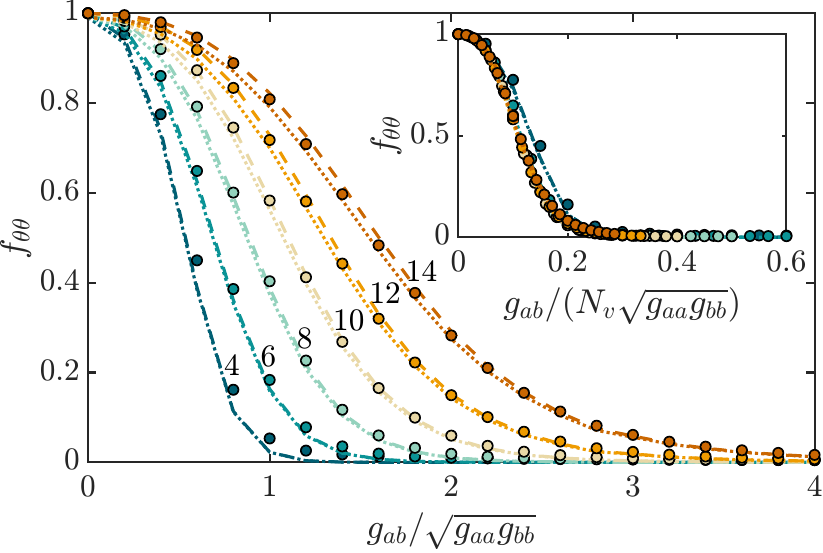}
    \caption{Superfluid fraction of component $b$ as a function of the intercomponent coupling $g_{ab}$, for different values of $N_v$ (black labels). Coloured dots correspond to the superfluid fraction $f_{\theta\theta}$ [Eq.~(\ref{eq:f_thetatheta})], while dotted and dashed lines correspond to the Leggett Bounds $f_{\theta\theta}^-$ and $f_{\theta\theta}^+$ defined in Eq.~(\ref{eq:f_thetatheta_+-}). Inset: same data, shown rescaling the horizontal axis by a factor $N_v$.} 
    \label{fig:Superfluid_fraction_various_N_v}
\end{figure}

To remain within linear response and compute $f_{\theta\theta}$ at the smallest possible $\Omega$, we choose a circulation along the inner ring given by $\mathfrak{n}_1 = -N_v/2$ and we compute the superfluid fraction defined in Eq.~(\ref{eq:f_thetatheta}) determining (by imaginary-time propagation) the stationary solutions of the system (of the type illustrated in Fig.~\ref{fig:Introductory_figure}) in a reference frame rotating with frequency $\Omega$ given by Eq.~\eqref{eq:Omega} [this requires adding an additional term $-\Omega \hat{L}_z\psi_i$ in each of the two Eqs.~\eqref{eq:GPEs}].  
For concreteness, we consider a Bose-Bose mixture of $^{23}\mathrm{Na}$ and $^{39}\mathrm{K}$ atoms, representing components $a$ and $b$ respectively, having intracomponent scattering lengths $a_{aa}=52\,a_0$ and $a_{bb}=8\,a_0$, with $a_0$ the Bohr radius~\cite{Schulze2018,Richaud2019}. This mixture is particularly suited to implement the discussed protocol, as the intercomponent scattering length $a_{ab}$ can be widely tuned via a Feshbach resonance across the miscible-immiscible transition~\cite{Richaud2019}. We further choose experimentally-relevant values $N_a\approx 2\times 10^5$ and $N_b\approx 5\times 10^3$, $d_z\approx 0.5\,\mu$m, $R_1=20\,\mu$m and $R_2=50\,\mu$m, leading to $r_0\approx 32\,\mu$m and $|\Omega| \approx 1.4\,\mathrm{rad/s}$. The resulting superfluid fractions are shown by colored dots in Fig.~\ref{fig:Superfluid_fraction_various_N_v}, where they are plotted as a function of the miscibility parameter $g_{ab}/\sqrt{g_{aa}g_{bb}}$ for various vortex numbers $N_v$.

A powerful method for estimating the degree of superfluidity of a given system is provided by the so-called Leggett's bounds~\cite{Leggett1970,Leggett1998}. Quite remarkably, these provide lower and upper limits for the superfluid fraction based solely on the density distribution $|\psi_b|^2$.
The upper bound  $f^+$ is derived variationally and is therefore rigorous, meaning that the actual superfluid fraction cannot exceed this limit. In contrast, the lower bound $f^-$ is heuristic and generally holds only for dilute Bose gases, where the system remains fully condensed~\cite{Chauveau2023,Perez2025}.
A growing number of studies have benchmarked these bounds against established measurements of the superfluid fraction~\cite{Chauveau2023,Tao2023,Pezze2024, Biagioni2024,Perez2025}, highlighting their limitations across different configurations~\cite{Orso2024,Perez2025}. When applied to the azimuthal response, these may be computed as
\begin{equation}
    f^-_{\theta\theta} =\frac{4\pi^2}{N_b}\int \frac{r\, \mathrm{d}r}{\int \frac{\mathrm{d}\theta}{|\psi_b|^2}}\quad{\rm and}\quad
    f^+_{\theta\theta} = \frac{4\pi^2}{N_b}\frac{1}{\int\frac{\mathrm{d}\theta}{\int r\,\mathrm{d}r \,|\psi_b|^2}}.
\label{eq:f_thetatheta_+-}
\end{equation} 
The values of the upper and lower bounds which we extract from the densities of component $b$ computed in imaginary time are illustrated in Fig.~\ref{fig:Superfluid_fraction_various_N_v} (dashed and dotted lines). Notably, the condition $f_{\theta\theta}^- \leq f_{\theta\theta} \leq f_{\theta\theta}^+$ is always satisfied,
confirming that the superfluid fraction of a dilute Bose gas (which is an intrinsically dynamic property) can be reliably extracted from its density profile only.

\begin{figure}[h!]
    \centering
    \includegraphics[width=\columnwidth]{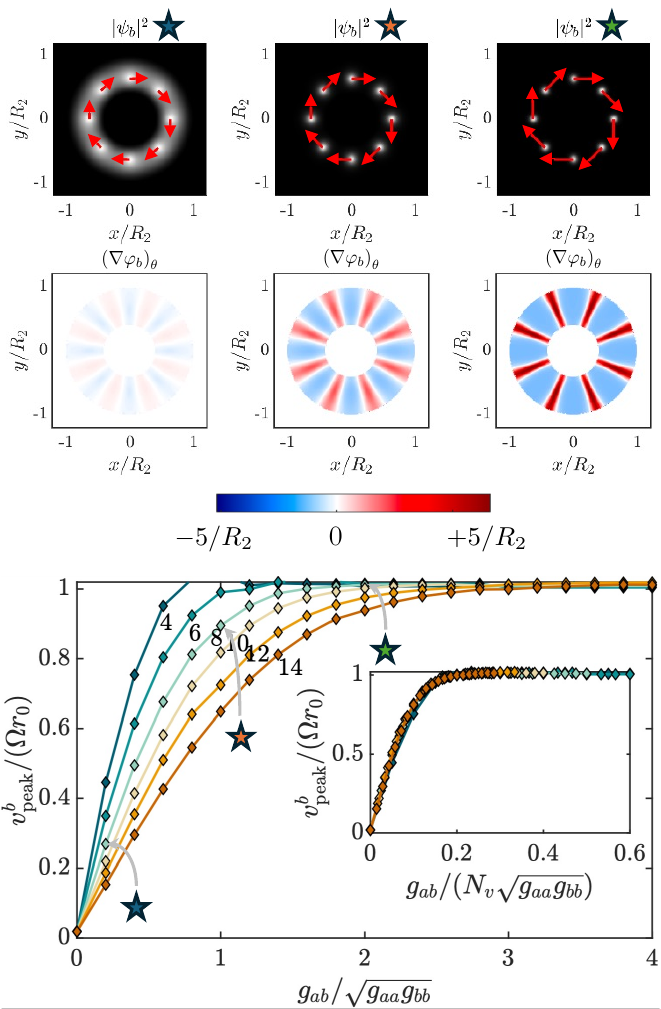}
    \caption{\textbf{Upper row:} Density of component $b$ for three different values of $g_{ab}$; red arrows denote the relevant velocity field evaluated at the density peaks. \textbf{Second row:} azimuthal gradient of the relevant phase field. \textbf{Lower panel:} peak velocity~(\ref{eq:v_peak_b}) as a function of $g_{ab}$ for different values of $N_v$ (black labels). Coloured stars serve as a visual guide and correspond to $g_{ab}/\sqrt{g_{aa}g_{bb}}=\{0.2,\,1.0,\,2.0\}$. Inset: data collapse on the same curve upon rescaling the horizontal axis by a factor $N_v$.}
\label{fig:Peak_velocity_various_N_v_assembled}
\end{figure}

\smallskip
\textit{Interaction-driven localization and drop of superfluidity.}
When  $g_{ab} = 0$, the two quantum fluids are fully decoupled. In this case, $\psi_b$  remains uniform along the $\theta$-direction, carrying zero angular momentum. Consequently, the superfluid fraction computed according to Eq.~(\ref{eq:f_thetatheta}) is exactly one. This result is consistent with the Leggett bounds, which also yield  $f_{\theta\theta}^+ = f_{\theta\theta}^- = 1$ as the density distribution $|\psi_b(r,\,\theta)|^2$  is independent of  $\theta$. As  $g_{ab}$  increases, component $b$ becomes more and more confined within the vortex cores of component $a$ (see upper row of Fig.~\ref{fig:Peak_velocity_various_N_v_assembled}). 
Consequently, the density profile $|\psi_b|^2$  acquires a stronger angular dependence, and the superfluid fraction decreases. This trend is consistently captured both by direct calculations using Eq.~(\ref{eq:f_thetatheta}) and by the Leggett bounds. In the deeply immiscible regime ($g_{ab} \gg \sqrt{g_{aa} g_{bb}}$), the density modulations become so strong that component $b$ effectively forms a necklace of spatially disconnected density peaks. 
In this insulating state $f_{\theta\theta}$ vanishes, indicating a complete loss of superfluid transport across the system. This behavior represents the precursor of the celebrated Mott transition~\cite{Fisher1989,Jaksch1998,Greiner2002,Chauveau2023}. Notably, in this regime, the vortices in component $a$ effectively bind to massive cores, giving place to massive quantum vortices, a class of dressed topological excitations currently under intense theoretical scrutiny~\cite{Richaud2020,Richaud2021PRA,Richaud2022,Richaud2023,Chaika2023,Patrick2023,Caldara2023,Doran2024,Bellettini2024,Dambroise2025,Caldara2024}. A related superfluid-to-insulator transition has been recently discussed in the context of dipolar gases, where rotational symmetry becomes spontaneously broken in the supersolid and droplet phases~\cite{Blakie2020,Smith2023,Chergui2023,Mukherjee2023,Sindik2024}.

The superfluid fraction further depends on the number of vortices in the necklace. As $N_v$ increases, the spacing between adjacent density peaks in $|\psi_b|^2$  decreases, leading to a greater overlap of their tails. This enhanced overlap facilitates phase coherence between neighboring density peaks, effectively strengthening superfluid behavior. As a result, higher-$N_v$  configurations transition to the Mott-like regime at larger values of  $g_{ab}$ compared to their smaller-$N_v$ counterparts (see also the inset).

The scenario described above is further confirmed by the analysis of an alternative indicator: the velocity
\begin{equation}
\label{eq:v_peak_b}
v_\mathrm{peak}^b = \frac{\hbar}{m_b} \left |\frac{1}{r} \frac{\partial\varphi_b}{\partial\theta}\right|_{\bm{r}_*}
\end{equation}
of the density peaks of component $b$ (where  $\varphi_b$ is the phase of the wavefunction $\psi_b$ and $\bm{r}_*$ is the position of a peak in $|\psi_b|^2$). As illustrated in the first row of Fig.~\ref{fig:Peak_velocity_various_N_v_assembled} (see, in particular, the red arrows), the tangential velocity of these peaks increases together with the loss of superfluid fraction, i.e., as component-$b$  atoms become increasingly localized within the vortex cores. In the weak-coupling limit ($g_{ab} \to 0$), the tangential velocity of the peaks tends to zero. In this regime, the underlying rotating potential imposed by component $a$ induces only local compressions and expansions in component $b$, without generating a significant net tangential flow.

Conversely, in the strongly immiscible regime ($g_{ab} \gg \sqrt{g_{aa} g_{bb}}$) component-$b$ atoms fully localize in the vortex cores, which behave as point-like particles. In this regime, they exhibit uniform circular motion with a tangential velocity given by $\Omega r_0$. Effectively, component $b$ forms a rigid necklace, where density peaks move collectively in a manner dictated by classical rotational dynamics. This crossover from a superfluid state to a state dominated by rigid-body motion is illustrated in the lower panel of Fig.~\ref{fig:Peak_velocity_various_N_v_assembled}. As previously discussed for $f_{\theta\theta}$, vortex necklaces with a larger number of vortices are more resilient to the onset of Mott-like localization and thus retain a significant superfluid fraction even at moderately large values of the miscibility parameter (see inset). 

To further characterize this crossover, we compute the azimuthal phase gradient $(\nabla \varphi_b)_\theta=r^{-1}\partial_\theta\varphi_b$. This quantity is illustrated in the middle row of Fig.~\ref{fig:Peak_velocity_various_N_v_assembled}, showing that its magnitude increases as  $g_{ab}$ increases. Moreover, one can observe that this phase gradient exhibits opposite signs in different regions: it is negative at the density peaks, reflecting their clockwise motion, while it is positive in the low-density regions between peaks. This structure originates from the irrotational nature of the superfluid velocity field and has been previously reported in different contexts, including rotating supersolids~\cite{Roccuzzo2020}.

\smallskip
\textit{Conclusions. --- } 
We have presented a platform for measuring the superfluid fraction of an atomic Bose-Einstein condensate without external perturbations or imposed optical lattices. Our method exploits the intrinsic rotation of vortex necklaces~\cite{Havelock1931,Fetter1967,Caldara2023,Caldara2024} in an auxiliary quantum gas, the component $a$, which naturally provides an effective periodic potential and an angular-momentum transfer to the one under test, the component $b$. This enables a direct measurement of the effective moment of inertia of $b$, and therefore of its superfluid fraction~\cite{Leggett1999,Stringari1996,Sindik2024}. We have illustrated this approach by analyzing a crossover from a fully superfluid condensate to a strongly localized state, where component $b$ forms an effectively classical array of density peaks, collectively moving as a rigid structure. 

A key result is that the extracted superfluid fraction is excellently bracketed by the Leggett bounds~\cite{Leggett1970,Leggett1998}, confirming both the validity of our protocol and the applicability of Leggett’s theory in this setting. Unlike conventional optical-lattice-based approaches, our system provides a self-generated, interaction-driven density modulation, eliminating the need for externally imposed lattices.

Beyond its equilibrium applications, our platform exhibits two key features that make it especially well-suited for investigating nonequilibrium phenomena, as detailed in the End Matter. First, the intrinsic dynamical instability of vortex necklaces~\cite{Baggaley2018,Hernandez2024,Giacomelli2023,Caldara2024} offers a natural avenue to study superfluidity in dynamically evolving, disordered landscapes. Second, the possibility to dynamically tune or quench the intercomponent interaction provides a controlled means to probe the time-dependent behavior of the superfluid fraction, an aspect that remains largely unexplored.

While our analysis focused on the linear-response regime, extending the study to larger rotation frequencies offers a compelling connection to the physics of finite-circulation states in Josephson junction arrays~\cite{Pezze2024}. These systems provide powerful testbeds for quantum circuits, where the interplay of macroscopic phase coherence, nonlinear effects, and dissipation governs key functionalities in quantum computing, simulation, and metrology. In this context, our platform opens promising avenues for future developments in quantum transport, atomtronics, and superfluid circuits~\cite{AmicoAVS2021,Amico2022,Polo2024}.

\vspace{5mm}

\begin{acknowledgments}
\textit{Acknowledgments. ---} A.R. received funding from the European Union’s Horizon research and innovation programme under the Marie Skłodowska-Curie grant agreement \textit{Vortexons} (no.~101062887). A.R. and P.M. acknowledge further support by the Spanish Ministerio de Ciencia e Innovación (MCIN/AEI/10.13039/501100011033, grant PID2023-147469NB-C21), by the Generalitat de Catalunya (grant 2021 SGR 01411), and by the ICREA {\it Academia} program. 
\end{acknowledgments}

\clearpage

\section*{End Matter}

\subsection{From regular to disordered potentials} 
Vortex necklaces are well known to be dynamically unstable. This instability is the superfluid counterpart~\cite{Baggaley2018,Giacomelli2023,Hernandez2024,Caldara2024} of the well-known Kelvin-Helmholtz instability (KHI), extensively studied in classical fluids~\cite{Landau1987}.
In the superfluid case, the characteristic breakdown timescale is given by~\cite{Aref1995,Caldara2024}
\begin{equation}
\label{eq:tau_star}
\tau^* = \frac{8 m_a r_0^2}{\hbar N_v^2 },
\end{equation}
which, in our system, is $\approx 0.056 \, \text{s}$.  The breakdown of the vortex necklace corresponds to a qualitative change in the structure of the effective potential $g_{ab}|\psi_a|^2$ felt by $\psi_b$: it evolves from a regular array of potential wells into a disordered landscape. 
Dilute Bose gases in disordered potentials attracted considerable attention in recent years~\cite{Fallani2007,Zuniga2015,Giorgini1994,Singh1994,Yukalov2007,Gaul2009,Lugan2011,Perez2025,Geier2025}, and our set-up permits to shine new light on their superfluid response.

\begin{figure}[t!]
    \centering
    \includegraphics[width=0.85\columnwidth]{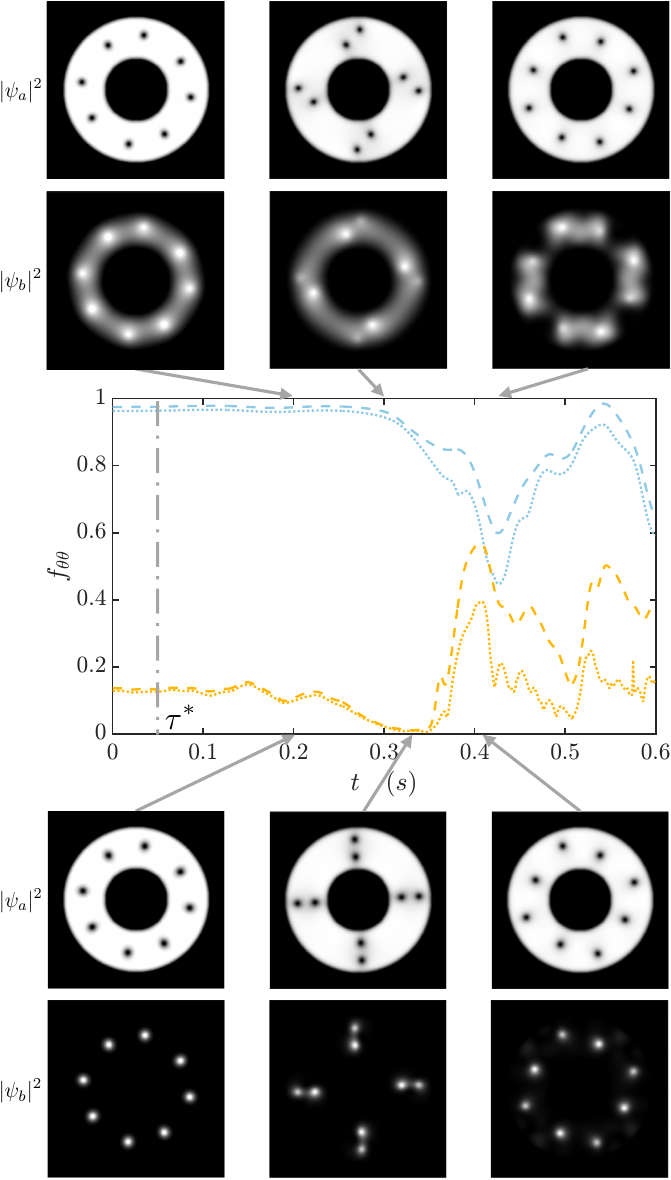}
    \caption{\textbf{Upper panels:} Breakdown and reformation of a regular vortex necklace in component $a$, accompanied by a significant change in the azimuthal density distribution of component $b$. Parameters detailed in the main text, here with $g_{ab}/\sqrt{g_{aa}g_{bb}}=0.20$. \textbf{Central panels:} Time evolution of the upper (dashed lines) and lower (dotted lines) Leggett bounds. The gray dash-dotted line corresponds to the timescale~(\ref{eq:tau_star}). \textbf{Lower panels:} Same as ``Upper panels", but with $g_{ab}/\sqrt{g_{aa}g_{bb}}=1.20$.}
    \label{fig:KHI_and_fs}
\end{figure}

Leggett’s bounds are derived for stationary systems, but one expects that they can still be applied in time-evolving scenarios provided that the dynamics is slow compared to all intrinsic timescales of the system. 
As discussed in Ref.~\cite{Chauveau2023}, the relevant energy scales here are:  
the depth of the effective lattice potential created by the vortex cores of component $a$, $\epsilon_1 = g_{ab} \bar{n}_a$; the recoil energy $\epsilon_2 = \hbar^2 k^2 / (2m_b)$, associated with the modulation wavenumber $k = N_v / r_0$; and the bulk chemical potential $\epsilon_3 = g_{bb} \bar{n}_b$, set by the average density $\bar{n}_b = N_b/[\pi(R_2^2 - R_1^2)]$.
When the KHI timescale  $\tau^*$ is larger than the three other characteristic timescales $\tau_i=\hbar/\epsilon_i$ (with $i=1,\,2,\,3$), the system should remain in quasi-equilibrium at each instant and Leggett’s bounds can be applied instantaneously, providing real-time upper and lower estimates for the superfluid fraction $f_{\theta\theta}$ during the disordering process. 
In our system $\tau_1 = 0.001$ s, $\tau_2 = 0.024$ s and $\tau_3 = 0.2$ s, so that $\tau_1<\tau_2<\tau^*<\tau_3$; but (as visible in Fig.~\ref{fig:KHI_and_fs}) the vortex necklace is still relatively stable at $\tau_3$; for this reason, the Leggett's bounds (dotted and dashed lines in Fig.~\ref{fig:KHI_and_fs}) are still expected to provide meaningful time-dependent bounds to $f_{\theta\theta}$. 

The outcome of a typical Gross-Pitaevskii simulation involving this process is shown in Fig.~\ref{fig:KHI_and_fs}, where we compare two different systems, one initially almost fully superfluid (upper panels and blue lines of the central panel), the other initially rather localized (lower panels and yellow lines of the central panel). 
The results reveal qualitatively and quantitatively distinct behaviors in response to the breakdown and (transient) reformation of the vortex necklace in component $a$. In the first system, which is deep in the miscible regime ($g_{ab}/\sqrt{g_{aa}g_{bb}}=0.20$), the superfluid fraction undergoes a significant drop, unexpectedly reaching its lowest value when a quasi-regular vortex necklace reforms ($t\approx 0.42\, s$). When this happens, in fact, component $b$ exhibits a sort of ``dimerization", that means that the relevant density distribution $|\psi_b|^2$ features a staggered pattern with respect to the $N_v$ ``bonds". In particular, the low-density regions associated with the $N_v/2$ weakly-populated ``bonds" disconnect high-density regions~\cite{Perez2025} and therefore are responsible for a sharp reduction of the superfluid fraction.      

The second system, being in the immiscible regime ($g_{ab}/\sqrt{g_{aa}g_{bb}}=1.20$), is characterized by a stronger confinement of component-$b$ bosons within component-$a$ vortex cores. Here, the lowest superfluid fraction ($f_{\theta\theta}\approx 0$) occurs when the vortices become pairwise radially aligned ($t\approx 0.34\, s$). This hinders tunneling in the azimuthal direction, leading to near-complete localization. However, in contrast to the former case, when the vortex necklace reforms ($t=0.42\,s$), the superfluid fraction recovers. This is due to the shaking induced by the underlying vortex motion, which causes component-$b$ atoms to escape from the relevant vortex cores, repopulating the inter-vortex regions and thus forming ``conducting channels"~\cite{Perez2025} that support superfluid flow.

\subsection*{Quenching the intercomponent interaction}
The platform introduced in this work is particularly well-suited to investigate nonequilibrium phenomena. An intriguing example is represented by the possibility of controllably ramping the intercomponent interaction $g_{ab}$, and hence to drive the system across the miscible-to-immiscible crossover~\cite{Schulze2018,Richaud2019}. To characterize the response of the system to this driving, and in particular the time-evolution of the superfluid fraction of component $b$, we define the ramp rate
\begin{equation}
\label{eq:gamma}
    \gamma := \frac{\mathrm{d}}{\mathrm{d}t} \frac{g_{ab}}{\sqrt{g_{aa}g_{bb}}}.
\end{equation}
We consider ramps whose total duration $\Delta t$ is shorter than the characteristic time associated to the onset of the Kelvin-Helmholtz instability in component $a$. Under this condition the vortex necklace remains structurally stable throughout the ramp, ensuring that component $b$ is subject to a quasi-stationary effective potential. 

\begin{figure}[b!]
    \centering
    \includegraphics[width=1\columnwidth]{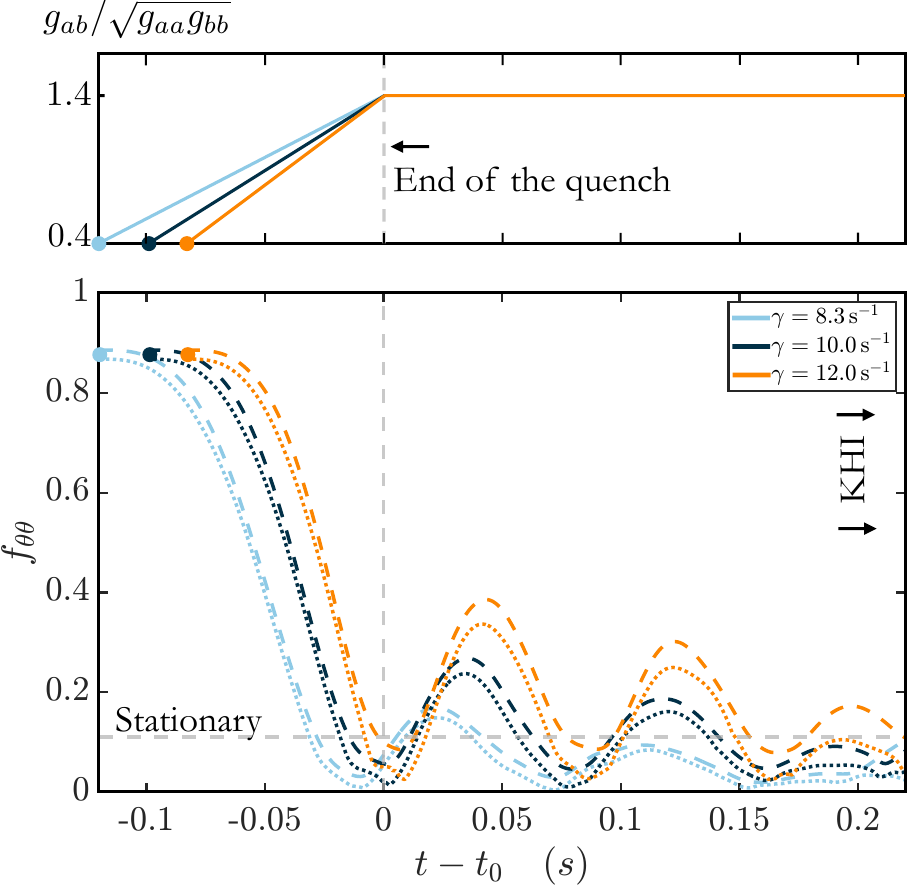}
    \caption{\textbf{Upper panel:} Quench protocol used for the intercomponent interaction $g_{ab}(t)$. After linearly increasing the ratio $g_{ab}/\sqrt{g_{aa} g_{bb}}$ at a rate $\gamma$, the system for $t\ge t_0$ is evolved according to the final constant set of parameters. \textbf{Lower panel:} upper (dashed lines) and lower (dotted lines) Leggett bounds  as a function of time during and following the quench from $g_{ab}/\sqrt{g_{aa} g_{bb}} = 0.4$ to $g_{ab}/\sqrt{g_{aa} g_{bb}} = 1.4$ in a system of $N_v=8$ vortices, for three different ramp rates $\gamma$. The horizontal gray line corresponds to the value of $f^+\approx f^-\approx 0.1$ computed in the stationary state (see Fig.~\ref{fig:Superfluid_fraction_various_N_v}).  }
    \label{fig:Kibble_Zurek}
\end{figure}

We implement the following protocol: the system is initially prepared in the miscible regime, i.e. with a value of the miscibility ratio $g_{ab}/\sqrt{g_{aa}g_{bb}}=0.4$ corresponding to a large superfluid fraction ($f_{\theta\theta}\approx 0.9$) in component $b$. The intercomponent interaction is then ramped up linearly at a rate $\gamma$ during a time $\Delta t$ to a final value $g_{ab}/\sqrt{g_{aa}g_{bb}}=1.4$ (hence in the immiscible regime), which is reached at time $t=t_0$, after which the interaction strength is held constant (see upper panel of Fig.~\ref{fig:Kibble_Zurek}). During and after the ramp, we track the evolution of the Leggett bounds in real time to monitor the instantaneous superfluid fraction (see lower panel of Fig.~\ref{fig:Kibble_Zurek}).

We observe that during the ramp both the upper and lower Leggett bounds decrease monotonically, reflecting the increasing degree of localization of component-$b$ atoms within the vortex cores. Once the ramp is complete and the interaction is held constant, i.e. for $t>t_0$, the Leggett bounds exhibit some (damped) oscillations, with two notable features: 
\textit{i)} They share the same frequency irrespective of the value of the ramp rate $\gamma$;
\textit{ii)} Their amplitude increases with $\gamma$.
The presence of these coherent post-quench oscillations can be understood in the framework of the Kibble-Zurek mechanism~\cite{Kibble1980,Laguna1997,Laguna1998},
 which explains how universal dynamical responses emerge after non-adiabatic ramps across a phase crossover.

Furthermore, the oscillations are centered around an average value ($\bar{f}_{\theta\theta}\approx 0.1$) which corresponds to the expected \emph{equilibrium} superfluid fraction for the final interaction strength (see horizontal gray line in Fig.~\ref{fig:Kibble_Zurek}). One can also verify that the periodic vanishing of the superfluid fraction is due to corresponding vanishing of the particle density in the inter-peak regions. 

Finally, the oscillations persist only as long as the vortex necklace remains stable. Once the Kelvin-Helmholtz instability sets in, the regularity of the effective potential gets lost, and the Leggett bounds cease to oscillate. 
A related nonequilibrium numerical experiment was presented by Kohn {\it et al.} in Ref.~\cite{Kohn2020}, who investigated the quench of bosons across the superfluid-Mott insulator transition in a rotating annular optical lattice.

Our results suggest that the vortex-necklace platform not only enables superfluid fraction measurements in quasi-static settings, but also offers a simple and tunable experimental playground for exploring the nonequilibrium dynamics of superfluid currents, with promising applications in atomtronics and quantum transport~\cite{AmicoAVS2021,Amico2022}. 

\end{document}